\documentclass{article}
\makeatletter

\@addtoreset{equation}{section}
\makeatother

\usepackage{amssymb}
\usepackage{amscd}
\usepackage{mathrsfs}
\usepackage{bm}
\usepackage{cite}

\bmdefine{\bx}{x}
\bmdefine{\by}{y}
\bmdefine{\bk}{k}
\bmdefine{\bp}{p}
\bmdefine{\bq}{q}

\begin{document}

\begin{titlepage}
\setcounter{page}{0}
\begin{flushright}
\begin{tabular}{l}
	{\tt arxiv:0709.1010}\\
	{\tt KEK-TH-1177}\\
\end{tabular}
\end{flushright}
\vskip 2.0cm

\begin{center}

	{\Large\bf On the Correspondence between \\
	Poincar\'e Symmetry of Commutative QFT and \\
	Twisted Poincar\'e Symmetry of Noncommutative QFT}

\vspace{15mm}

\renewcommand{\thefootnote}{\fnsymbol{footnote}}
	{\large Yasumi Abe\footnote[1]{email:\ yasumi@post.kek.jp}}\\[10mm]
	\noindent {\em Institute of Particle and Nuclear Studies\\
	High Energy Accelerator Research Organization (KEK)\\
	Tsukuba 305-0801, Japan }\\
\vspace{20mm}

\begin{abstract}
	The space-time symmetry of noncommutative quantum field theories with a deformed quantization is described by the twisted Poincar\'e algebra, while that of standard commutative quantum field theories is described by the Poincar\'e algebra.
	Based on the equivalence of the deformed theory with a commutative field theory, the correspondence between the twisted Poincar\'e symmetry of the deformed theory and the Poincar\'e symmetry of a commutative theory is established.
	As a by-product, we obtain the conserved charge associated with the twisted Poincar\'e transformation to make the twisted Poincar\'e symmetry evident in the deformed theory.
	Our result implies that the equivalence between the commutative theory and the deformed theory holds in a deeper level, i.e., it holds not only in correlation functions but also in (different types of) symmetries.
\end{abstract}

\end{center}
\end{titlepage}

\section{Introduction}\mbox{}

	The twisted Poincar\'e algebra is a quantum group that is obtained by Drinfel'd twist of the universal enveloping algebra $\mathcal{U}(\mathcal{P})$ of the Poincar\'e algebra $\mathcal{P}$.
	It describes the symmetry of noncommutative space-time whose coordinates obey the commutation relation of a canonical type,
\begin{equation}
	[\hat{x}^\mu,\hat{x}^\nu]=i\theta^{\mu\nu}.
\end{equation}
	The twisted Poincar\'e symmetry has been proposed in \cite{Chaichian:2004za} as a substitute for the Poincar\'e symmetry in field theories on the noncommutative space-time.
	In terms of the twisted Poincar\'e symmetry, the Moyal star product
\begin{equation}
	f(x)\ast g(x)=\exp\left[\frac{i}{2}\theta^{\mu\nu}\partial_\mu'\partial_\nu''\right]f(x')g(x'')\Big\vert_{x',x''\rightarrow x},
\end{equation}
	which provides the noncommutative product for fields on the noncommutative space-time is obtained as a twisted product of a module algebra of the twisted Poincar\'e algebra.
	This fact implies the twisted Poincar\'e invariance of noncommutative field theories.

	Recently, some researchers including the author have proposed a quantum field theory (QFT) which possesses the twisted Poincar\'e symmetry\cite{Chaichian:2004yh,Balachandran:2005eb,Balachandran:2005pn,Lizzi:2006xi,Bu:2006ha,Abe:2006ig,Fiore:2007vg}.
	In this QFT, the star product on different space-time points is used as a product for fields, and thus it can be considered as a deformed theory of a standard commutative QFT.
	Taking account of the role of the Poincar\'e symmetry played for the standard commutative QFT, it seems worthwhile to investigate the consequences such a deformation yields thoroughly.
	Clarification of the property of the theory associated to the twisted Poincar\'e symmetry may lead to a fuller understanding of the implication of the noncommutativity for quantum field theories.

	The deformation through the star product brings two remarkable properties to the new QFT.
	One is the twisted Poincar\'e invariance as mentioned above.
	The other is that correlation functions of the deformed QFT appear to take the same values as those of the corresponding commutative QFT\footnote{The equivalence of correlation functions holds depending on the definition of correlation functions in the deformed theory. In \cite{Oeckl:2000eg}, correlation functions are defined without the star product. Constructing the deformed QFT based on this correlation function, one find the resulting dynamics to be different from that of a commutative QFT. For speculation on the Hopf algebraic symmetry of this theory, see \cite{Oeckl:1999zu,Sasai:2007me}.}.
	In fact, one can construct a map between field operators of the two theories which suggests the equivalence of correlation functions\cite{Abe:2006ig}.
	It is noticed, however, that this equivalence has not been verified rigorously as we shall explain in section 2.
	In this paper, we assume this equivalence and investigate a consequence of it.
	Once the equivalence is admitted, it implies, in some sense, a discouraging fact that the nontrivial deformation of the theory results in no new dynamics: the dynamics of the new QFT is exactly the same as that of the commutative QFT.
	On the other hand, it means that any troublesome properties inherent in the ordinary noncommutative QFT, such as UV/IR mixing \cite{Minwalla:1999px}, disappear in the new deformed QFT, and one can obtain a well defined QFT as long as the corresponding commutative QFT is well defined.

	Now, what does this equivalence imply for symmetries?
	From the fact that the deformed QFT is twisted Poincar\'e covariant while the commutative QFT is Poincar\'e covariant, it is expected that these two different symmetries correspond with each other through the equivalence of the two theories, that is, the Poincar\'e transformations in commutative QFTs may be represented as the corresponding twisted Poincar\'e transformations in deformed QFTs.
	The purpose of this paper is to show that this is indeed the case with the statement presented as a theorem.
	To this end, we use the map between the two theories presented in \cite{Abe:2006ig}, and thereby obtain generators of Poincar\'e transformations in the deformed QFT from those in the commutative QFT.
	The twisted Poincar\'e symmetry of the deformed QFT can then be derived by twisting the Poincare algebra constructed from these generators.
	For definiteness, we will restrict our attention to a real scalar field in $d+1$ dimensional Minkowski space-time with metric $(+,-,\cdots,-)$ whose interaction term is given by polynomials.
	Further we assume that the time and space coordinates are commutative with each other, i.e., $\theta^{0i}=0$, so that the discussion of noncommutative field theories in terms of a canonical formalism can be presented in a simple form.
	Presumably, this assumption is not essential to results presented here \cite{Fiore:2007vg}.

	This paper is organized as follows.
	In section 2, we recall the main result of \cite{Abe:2006ig} which is needed for our discussion.
	Section 3 is devoted to the investigation of the twisted Poincar\'e invariance of the deformed QFT.
	We present the standard Poincar\'e invariance of commutative QFTs in terms of the Hopf algebraic structure of $\mathcal{U}(\mathcal{P})$ in section 3.1.
	The Poincar\'e algebra represented in the commutative QFT is translated to that represented in the deformed QFT by the map between the deformed QFT and the commutative QFT.
	Then the Poincar\'e algebra in the deformed QFT is twisted in order to describe the symmetry of the deformed QFT.
	With these preparations, we provide the proof of the equivalence between a twisted Poincar\'e transformation in the deformed QFT and a Poincar\'e transformation in the commutative QFT in section 4.
	Our conclusions and remarks are given in section 5.

\section{Noncommutative field theory with deformed quantization}\mbox{}

	The deformed QFT with twisted Poincar\'e symmetry has been investigated in \cite{Chaichian:2004yh,Balachandran:2005eb,Balachandran:2005pn,Lizzi:2006xi,Bu:2006ha,Abe:2006ig,Fiore:2007vg}.
	There are some different approaches to define this theory.
	In \cite{Abe:2006ig}, we have taken a star product of fields at different space-time points to define the deformed QFT:\begin{equation}
	f(x)\star g(y):=\exp\left[\frac{i}{2}\partial^x\theta\partial^y\right]f(x)g(y)\label{star product}
\end{equation}
	where we introduce the notation $\partial^x\theta\partial^y:=\partial_{x^i}\theta^{ij}\partial_{y^j}$, which will be used generally for a contraction, $p\theta k:=p_i\theta^{ij}k_j$.
	By using this star product, we have seen that we can construct a well defined quantum field theory, by starting from the following Lagrangian
\begin{equation}
	\mathcal{L}^\theta(x)=\frac{1}{2}\Big[(\partial_\mu\phi^\theta)^2-m^2(\phi^\theta)^2\Big]-\sum_{n=3}\frac{\lambda_n}{n!}\overbrace{\phi^\theta\star\cdots\star\phi^\theta}^n \label{noncommutative lagrangian}
\end{equation}
	and quantizing the field through a deformed commutation relation
\begin{equation}
\begin{array}{rcl}
	~[\phi^\theta(t,\bx),\pi^\theta(t,\by)]_\star&=&\phi^\theta(t,\bx)\star\pi^\theta(t,\by)-\pi^\theta(t,\by)\star\phi^\theta(t,\bx)\\
	&=&i\delta^{(d)}(\bx-\by)\\
	~[\phi^\theta(t,\bx),\phi^\theta(t,\by)]_\star&=&[\pi^\theta(t,\bx),\pi^\theta(t,\by)]_\star=0,\rule{0mm}{5mm}
\end{array}
	\label{deformed quantization}
\end{equation}
	where $\pi^\theta=\partial_0\phi^\theta$.
	We call the theory given by (\ref{noncommutative lagrangian}) and (\ref{deformed quantization}) as {\it a deformed noncommutative quantum field theory} (dNCQFT) in this paper.
	Let us define correlation functions of field operators between arbitrary states in a Hilbert space $\mathcal{H}^\theta$ which carries a representation of $\phi^\theta$ as\footnote{The definition of the star product between operators and states will be given in section 3.2.}
\begin{equation}
	\langle\alpha|\star\phi^\theta(x_1)\star\cdots\star\phi^\theta(x_n)\star|\beta\rangle,\quad |\alpha\rangle, |\beta\rangle\in\mathcal{H}^\theta.\label{star correlation function}
\end{equation}
	Then these correlation functions turn out to have the same value as those of a commutative quantum field theory (CQFT) whose Lagrangian is given by
\begin{equation}
	\mathcal{L}^0(x)=\frac{1}{2}\Big[(\partial_\mu\phi^0)^2-m^2(\phi^0)^2\Big]-\sum_{n=3}\frac{\lambda_n}{n!}(\phi^0)^n,
\end{equation}
	in which the field is quantized by the standard canonical commutation relation.
	That is, there is a correspondence between a state $|\alpha\rangle$ in $\mathcal{H}^\theta$ and a state $|\alpha'\rangle$ in $\mathcal{H}^0$ which carries a representation of $\phi^0$, and we have
\begin{equation}
	\langle\alpha|\star\phi^\theta(x_1)\star\cdots\star\phi^\theta(x_n)\star|\beta\rangle=\langle\alpha'|\phi^0(x_1)\cdots\phi^0(x_n)|\beta'\rangle.\label{equivalence of correlation}
\end{equation}
	This equivalence is found from the following map between $\phi^0$ and $\phi^\theta$:
\begin{equation}
\begin{array}{rcl}
	\phi^\theta(x)&=&{\displaystyle \exp\left[\frac{1}{2}\partial\theta P\right]\phi^0(x)}\\
	&=&{\displaystyle \sum_{n=0}^\infty\frac{1}{2^nn!}\theta^{i_1j_1}\cdots\theta^{i_nj_n}\partial_{i_1}\cdots\partial_{i_n}\phi^0(x)P_{j_1}\cdots P_{j_n}},\\
	\phi^0(x)&=&{\displaystyle \exp\left[-\frac{1}{2}\partial\theta P^\theta\right]\phi^\theta(x)}\\
	&=&{\displaystyle \sum_{n=0}^\infty\left(-\frac{1}{2}\right)^n\frac{1}{n!}\theta^{i_1j_1}\cdots\theta^{i_nj_n}\partial_{i_1}\cdots\partial_{i_n}\phi^\theta(x)P_{j_1}^\theta\cdots P_{j_n}^\theta},\label{map}
\end{array}
\end{equation}
	where $P_i$ and $P_i^\theta$ are generators of translations in CQFT and dNCQFT respectively\footnote{In fact, $P_i=P_i^\theta$ as we will see in the next section. From this relation, we confirm that the two equations in (\ref{map}) are in the relation of the inverse map with each other.}.
	In addition, based on this map, we use the same Hilbert space as a representation space of the field operator for both CQFT and dNCQFT, that is, we take $\mathcal{H}^\theta=\mathcal{H}^0$, and $|\alpha\rangle=|\alpha'\rangle$ and $|\beta\rangle=|\beta'\rangle$ in (\ref{equivalence of correlation}).
	In the following, we will denote this Hilbert space by $\mathcal{H}$.

	The correspondence between correlation functions (\ref{equivalence of correlation}) can be seen by noticing the following equation:
\begin{equation}
	O_\star(\phi^\theta)=\sum_n\left(\frac{i}{2}\right)^n\frac{1}{n!}\theta^{i_1j_1}\theta^{i_2j_2}\cdots\theta^{i_nj_n}[P_{i_1},[P_{i_2},\cdots[P_{i_n},O(\phi^0)]\cdots]]P_{j_1}P_{j_2}\cdots P_{j_n},\label{operator correspondence}
\end{equation}
	where $O(\phi^0)$ is an arbitrary operator constructed from $\phi^0$ by the ordinary product, and $O_\star(\phi^\theta)$ is an operator replacing all the fields and products between them in $O(\phi^0)$ by $\phi^0$ and the star product\footnote{Inversely, one may consider (\ref{operator correspondence}) as a definition of $O_\star(\phi^\theta)$ which corresponds to $O(\phi^0)$.}.
	For example, let us consider the case of $O(\phi^0)=\phi^0(x_0)\cdots\phi^0(x_n)$, in which the corresponding operator $O_\star(\phi^\theta)$ is given by $\phi^\theta(x_1)\star\cdots\star\phi^\theta(x_n)$.
	In this case, (\ref{operator correspondence}) reads
\begin{equation}
	\phi^\theta(x_1)\star\cdots\star\phi^\theta(x_n)=\exp\left[\frac{1}{2}\sum_n\partial_{x_n}\theta P\right]\phi^0(x_1)\cdots\phi^0(x_n),
\end{equation}
	and this is found by substituting the second equation of (\ref{map}) on the right hand side.
	Based on this equation, we can easily verify (\ref{equivalence of correlation}).

	Thus we ''prove" the equivalence of correlation functions of the two theories.
	However, it should be noted that this proof is somewhat formal, for we ignore some points which should be treated more carefully.
	Firstly, in the map between operators of the two theory (\ref{map}) or (\ref{operator correspondence}), the mapped operator is given by a nonlocal form of the original local operator, therefore we have to look into properties of the map, such as an asymptotic behavior, more carefully.
	Correspondingly, it is unclear whether we can take $\mathcal{H}^\theta=\mathcal{H}^0$ or not.
	Even if asymptotic completeness is satisfied for both theories, there would be no need for asymptotic states in them to correspond with each other in the simple way as we have stated above.
	There might be a representation, and thus asymptotic states peculiar to dNCQFT.
	This would spoil the equivalence of correlation functions.
	Though there would be need to examine the validity of this correspondence of the two theories more carefully, we assume it in this paper.
	In particular, we assume that asymptotic states behave in the same manner in both theories and $\mathcal{H}^\theta=\mathcal{H}^0$.
	It is noteworthy that, even without (\ref{map}) and (\ref{operator correspondence}), once $\mathcal{H}^\theta=\mathcal{H}^0$ is assumed, the equivalence of the correlation functions is proved in all order of perturbation\cite{Abe:2006ig}.

\section{Poincar\'e symmetry and twisted Poincar\'e symmetry}\mbox{}

	In this section, we show that dNCQFT has the twisted Poincar\'e symmetry.
	The twisted Poincar\'e symmetry of dNCQFT can be understood in terms of the Drinfel'd twist by $\mathcal{F}=e^{\frac{i}{2}\theta^{ij}P_i\otimes P_j}$.
	In dNCQFT, we can construct generators of Poincar\'e transformations from the field operator, therefore they form an algebra generated from the field operator on a representation space of them.
	By twisting the Poincar\'e algebra by $\mathcal{F}$, we obtain the twisted Poincar\'e algebra.
	Correlation functions of dNCQFT turn out to be invariant under a transformation in this twisted algebra.

\subsection{Poincar\'e symmetry of a commutative QFT and Hopf algebra}\mbox{}

	As a preliminary for introducing the twisted Poincar\'e symmetry of dNCQFT, we present the Poincar\'e symmetry of CQFTs in terms of the Hopf algebraic structure of $\mathcal{U}(\mathcal{P})$.
	Here $\mathcal{U}(\mathcal{P})$ is equipped with a coproduct $\Delta:\mathcal{U}(\mathcal{P})\rightarrow\mathcal{U}(\mathcal{P})\otimes\mathcal{U}(\mathcal{P})$, a counit $\varepsilon:\mathcal{U}(\mathcal{P})\rightarrow\mathbb{C}$ and an antipode $S:\mathcal{U}(\mathcal{P})\rightarrow\mathcal{U}(\mathcal{P})$ in addition to the algebraic structure as an enveloping algebra.
	These linear maps are given by standard definitions for an enveloping algebra of a Lie algebra.
	For precise definitions of them, see, for example, \cite{Chari:1994pz}.

	The Poincar\'e algebra $\mathcal{P}$, for which commutators of generators are given by
\begin{equation}
\begin{array}{l}
	~[P_\mu,P_\nu]=0,\\
	~[M_{\mu\nu},M_{\rho\sigma}]=-i(g_{\mu\rho}M_{\nu\sigma}-g_{\nu\rho}M_{\mu\sigma}-g_{\mu\sigma}M_{\nu\rho}+g_{\nu\sigma}M_{\mu\rho}),\rule{0mm}{5mm}\\
	~[M_{\mu\nu},P_\rho]=-i(g_{\mu\rho}P_\nu-g_{\nu\rho}P_\mu),\rule{0mm}{5mm}
\end{array}
	\label{poincare}
\end{equation}
	is represented in CQFT by
\begin{equation}
	P_\mu=\int d^dxT_{0\mu}(x),\quad M_{\mu\nu}=\int d^dx\left[x_\mu T_{0\nu}(x)-x_\nu T_{0\mu}(x)\right],\label{rep. of poincare}
\end{equation}
	where
\begin{equation}
	T_{0\mu}(x)=\frac{1}{2}\left(\pi^0(x)\partial_\mu\phi^0(x)+\partial_\mu\phi^0(x)\pi^0(x)\right)-g_{0\mu}\mathcal{L}^0(x),\label{rep. in CQFT}
\end{equation}
	and $\pi^0=\partial_0\phi^0$ is the canonical momentum of $\phi^0$.
	Of course, these operators are constant in time:
\begin{equation}
\begin{array}{l}
	{\displaystyle \frac{dP_\mu}{dt}=\frac{1}{i}[H^0,P_\mu]=0,}\\
	{\displaystyle \frac{dM_{\mu\nu}}{dt}=\frac{\partial M_{\mu\nu}}{\partial t}+\frac{1}{i}[H^0,M_{\mu\nu}]=0,}\rule{0mm}{7mm}
\end{array}
	\label{conservation}
\end{equation}
	where $H^0=P_0$.
	It is trivial to construct the representation of $\mathcal{U}(\mathcal{P})$ from this representation of $\mathcal{P}$.
	For the representation (\ref{rep. of poincare}) and (\ref{rep. in CQFT}), we can take the following two vector spaces as a representation space.

	One is the Hilbert space $\mathcal{H}$ (or its dual space $\mathcal{H}^*$) on which the field operator $\phi^0$ is represented.
	Denoting the action of $X\in\mathcal{U}(\mathcal{P})$ to $\mathcal{H}$ and $\mathcal{H}^*$ as
\begin{equation}
\begin{array}{ll}
	X(|\alpha\rangle)=X|\alpha\rangle,&|\alpha\rangle\in\mathcal{H},\\
	X(\langle\alpha|)=\langle\alpha|S(X),&\langle\alpha|\in\mathcal{H}^*,\rule{0mm}{5mm}
\end{array}
\end{equation}
	we can see that this action is compatible with the inner product of $\mathcal{H}$.
	That is, if we write the inner product of $\mathcal{H}$ by the pairing map ${\rm ev}:\mathcal{H}^*\otimes\mathcal{H}\rightarrow\mathbb{C}$,
\begin{equation}
	{\rm ev}(\langle\alpha|\otimes|\beta\rangle)=\langle\alpha|\beta\rangle,\label{inner product}
\end{equation}
	then
\begin{equation}
\begin{array}{lcl}
	X({\rm ev}(\langle\alpha|\otimes|\beta\rangle))&=&{\rm ev}(\Delta(X)(\langle\alpha|\otimes|\beta\rangle))\\
	&=&\langle\alpha|{\rm m}((S\otimes 1)\circ\Delta(X))|\beta\rangle=\langle\alpha|\varepsilon(X)|\beta\rangle,\rule{0mm}{5mm}
\end{array}
	\label{action to inner product}
\end{equation}
	where ${\rm m}:\mathcal{U}(\mathcal{P})\otimes\mathcal{U}(\mathcal{P})\rightarrow\mathcal{U}(\mathcal{P})$ is the product map of $\mathcal{U}(\mathcal{P})$ and we use the standard formula of a Hopf algebra,
\begin{equation}
	{\rm m}((S\otimes 1)\circ\Delta(X))=\varepsilon(X)\Big(={\rm m}((1\otimes S)\circ\Delta(X))\Big).\label{formula for antipode}
\end{equation}
	From the explicit value of the counit $\varepsilon$,
\begin{equation}
\begin{array}{ll}
	\varepsilon(c)=c,&c\in\mathbb{C}\subset\mathcal{U}(\mathcal{P}),\\
	\varepsilon(\chi)=0,&\textrm{$\chi\in\mathcal{P}\subset\mathcal{U}(\mathcal{P})$},\rule{0mm}{5mm}
\end{array}
	\label{counit}
\end{equation}
	we see that (\ref{action to inner product}) means the invariance of the inner product of $\mathcal{H}$ under a Poincar\'e transformation, since
\begin{equation}
\begin{array}{ll}
	c(\langle\alpha|\beta\rangle)=c\langle\alpha|\beta\rangle,&\textrm{for $c\in\mathbb{C}\subset\mathcal{U}(\mathcal{P})$,}\\
	\chi(\langle\alpha|\beta\rangle)=0,&\textrm{for $\chi\in\mathcal{P}\subset\mathcal{U}(\mathcal{P})$.}\rule{0mm}{5mm}\\
\end{array}
	\label{action to inner product 2}
\end{equation}
	It is clear that this implies the invariance of the inner product under an arbitrary transformation in $\mathcal{U}(\mathcal{P})$.

	The other representation space of $\mathcal{P}$ and $\mathcal{U}(\mathcal{P})$ is the algebra $\mathcal{M}(\phi^0)$ generated from the field operator $\phi^0$.
	The action of $P_\mu,M_{\mu\nu}\in\mathcal{P}$ on $\phi^0$ is given by the standard form:
\begin{equation}
\begin{array}{l}
	P_\mu(\phi^0):=[P_\mu,\phi^0]=-i\partial_\mu\phi^0,\\
	M_{\mu\nu}(\phi^0):=[M_{\mu\nu},\phi^0]=-i(x_\mu\partial_\nu-x_\nu\partial_\mu)\phi^0,\rule{0mm}{5mm}
\end{array}
	\label{action of poincare}
\end{equation}
	and the action of an arbitrary element of $\mathcal{U}(\mathcal{P})$ is obtained through $X_1X_2(\phi^0):=X_1(X_2(\phi^0))$, where $X_1,X_2\in\mathcal{U}(\mathcal{P})$.
	For example, the action of $P_{\mu_1}P_{\mu_2}\cdots P_{\mu_n}\in\mathcal{U}(\mathcal{P})$ on $\phi^0$ is
\begin{equation}
\begin{array}{rcl}
	P_{\mu_1}P_{\mu_2}\cdots P_{\mu_n}(\phi^0)&=&[P_{\mu_1},[P_{\mu_2},\cdots[P_{\mu_n},\phi^0]\cdots]]\\
	&=&(-i)^n\partial_{\mu_1}\cdots\partial_{\mu_n}\phi^0.\rule{0mm}{5mm}
\end{array}
\end{equation}
	Further, $\mathcal{M}(\phi^0)$ represents $\mathcal{U}(\mathcal{P})$ as a module algebra.
	In fact, denoting the product map of $\mathcal{M}(\phi^0)$ by $\mu:\mathcal{M}(\phi^0)\otimes\mathcal{M}(\phi^0)\rightarrow\mathcal{M}(\phi^0)$,
\begin{equation}
	\mu(O_1\otimes O_2)=O_1O_2,\quad\textrm{for $O_1, O_2\in\mathcal{M}(\phi^0)$,}
\end{equation}
	the action of $X\in\mathcal{U}(\mathcal{P})$ to the product is written as
\begin{equation}
	X(\mu(O_1\otimes O_2))=\mu(\Delta(X)(O_1\otimes O_2)).\label{action to product for operators}
\end{equation}

	Since $\mathcal{M}(\phi^0)$ is represented on $\mathcal{H}$ and $\mathcal{H}^*$, we can consider the compatibility between the action of $\mathcal{M}(\phi^0)$ to $\mathcal{H}$ and the action of $\mathcal{U}(\mathcal{P})$ to them.
	That is, writing the action of $\mathcal{M}(\phi^0)$ to $\mathcal{H}$ and $\mathcal{H}^*$ by linear maps $\mu_R:\mathcal{M}(\phi^0)\otimes\mathcal{H}\rightarrow\mathcal{H}$ and $\mu_L:\mathcal{H}^*\otimes\mathcal{M}(\phi^0)\rightarrow\mathcal{H}^*$ respectively as
\begin{equation}
\begin{array}{l}
	\mu_R(O\otimes|\alpha\rangle)=O|\alpha\rangle,\\
	\mu_L(\langle\alpha|\otimes O)=\langle\alpha|O,\rule{0mm}{5mm}
\end{array}
\end{equation}
	the action of $\mathcal{U}(\mathcal{P})$ to these states is written as
\begin{equation}
\begin{array}{l}
	X(\mu_R(O\otimes|\alpha\rangle))=\mu_R(\Delta(X)(O\otimes|\alpha\rangle)),\\
	X(\mu_L(\langle\alpha|\otimes O))=\mu_L(\Delta(X)(\langle\alpha|\otimes O)).\rule{0mm}{5mm}
\end{array}
	\label{action to state with operator}
\end{equation}
	In addition, we can introduce a linear map $\widetilde{\rm ev}:\mathcal{H}^*\otimes\mathcal{M}(\phi^0)\otimes\mathcal{H}\rightarrow\mathbb{C}$ for matrix elements of operators,
\begin{equation}
	\widetilde{\rm ev}(\langle\alpha|\otimes O\otimes |\beta\rangle)=\langle\alpha|O|\beta\rangle.
\end{equation}
	By composing ${\rm ev}$ with $\mu_R$ or $\mu_L$, $\widetilde{\rm ev}$ is rewritten as
\begin{equation}
	\widetilde{\rm ev}={\rm ev}\circ(1\otimes\mu_R)={\rm ev}\circ(\mu_L\otimes 1).\label{map for matrix element}
\end{equation}
	Using this expression, we can see the compatibility between $\widetilde{\rm ev}$ and the action of $\mathcal{M}(\phi^0)$:
\begin{equation}
\begin{array}{lcl}
	X\widetilde{\rm ev}(\langle\alpha|\otimes O\otimes |\beta\rangle)&=&\widetilde{\rm ev}((\Delta\otimes 1)\circ\Delta(X)(\langle\alpha|\otimes O\otimes |\beta\rangle))\\
	&=&\widetilde{\rm ev}((1\otimes\Delta)\circ\Delta(X)(\langle\alpha|\otimes O\otimes |\beta\rangle)).\rule{0mm}{5mm}
\end{array}
	\label{action to matrix element}
\end{equation}
	It is easily seen that (\ref{action to matrix element}) means the invariance of matrix elements of operators under a Poincar\'e transformation in the same way as (\ref{action to inner product 2}).
	By using the relation (\ref{map for matrix element}) and (\ref{formula for antipode}), (\ref{action to matrix element}) is written as
\begin{equation}
\begin{array}{lcl}
	X(\widetilde{\rm ev}(\langle\alpha|\otimes O\otimes|\beta\rangle))&=&\langle\alpha|\varepsilon(X)(O|\beta\rangle)\\
	&=&(\langle\alpha|O)\varepsilon(X)|\beta\rangle.\rule{0mm}{5mm}
\end{array}
	\label{invariance of matrix elements}
\end{equation}
	Again, from the explicit value of the counit $\varepsilon$, we see that this equation means the invariance of the matrix element under a Poincar\'e transformation.

	Finally, we notice that there hold some relations between linear maps introduced here, corresponding to the associativity of their action.
	For example,
\begin{equation}
	(O_1O_2)|\alpha\rangle=O_1(O_2|\alpha\rangle)\Leftrightarrow\mu_R\circ(\mu\otimes 1)=\mu_R\circ(1\otimes\mu_R).\label{associativity for operators and state}
\end{equation}

\subsection{Twisted Poincar\'e symmetry of a noncommutative QFT with the deformed quantization}\mbox{}

	To obtain the twisted Poincar\'e algebra $\mathcal{U}^\mathcal{F}(\mathcal{P})$ and a twisted module algebra of it, we start from the standard Poincar\'e algebra and its representation space, and then twist them.
	In dNCQFT, we can construct the Poincar\'e algebra by applying (\ref{map}) to (\ref{rep. in CQFT}) and substituting them in (\ref{rep. of poincare}).
	Then we acquire $P_\mu$ and $M_{\mu\nu}$ in the same form as (\ref{rep. of poincare}) but now $T_{0\mu}$ in it is given by
\begin{equation}
	T_{0\mu}=\sum_{n=0}^{\infty}\left(-\frac{1}{2}\right)^n\frac{1}{n!}\theta^{i_1j_1}\cdots\theta^{i_nj_n}\partial_{i_1}\cdots\partial_{i_n}\left[\frac{1}{2}\left(\pi^\theta\star\partial_\mu\phi^\theta+\partial_\mu\phi^\theta\star\pi^\theta\right)-g_{0\mu}\mathcal{L}^\theta\right]P_{j_1}^\theta\cdots P_{j_n}^\theta,\label{rep. in dNCQFT}
\end{equation}
	instead of (\ref{rep. in CQFT}).
	It is obvious that the resulting operators satisfy commutation relations of Poincar\'e algebra (\ref{poincare}).
	Since, to derive these operators, we only rewrite field operators in them according to (\ref{map}), their commutation relations do not change.
	Thus we can construct Poincar\'e algebra $\mathcal{P}$ and the universal enveloping algebra $\mathcal{U}(\mathcal{P})$ in dNCQFT.
	Notice that $P_\mu\in\mathcal{P}$ are equal to translation generators $P_\mu^\theta$ which are derived from Noether currents in terms of translations in dNCQFT.
	In fact, the difference between them is only total derivative terms in their integrand:
\begin{equation}
\begin{array}{lcl}
	P_\mu&=&{\displaystyle \int d^dxT_{0\mu}}\\
	&=&{\displaystyle \int d^dx \left[\frac{1}{2}\left(\pi^\theta\partial_\mu\phi^\theta+\partial_\mu\phi^\theta\pi^\theta\right)-g_{0\mu}\mathcal{L}^\theta+(\textrm{total derivative terms})\right]}.
\end{array}
\end{equation}
	Since we assume the correspondence of asymptotic behaviors of the two theories, this contribution does vanish to give $P_\mu=P_\mu^\theta$.
	In particular, the Hamiltonian $H^\theta=P_0^\theta$ in dNCQFT is equal to $H^0=P_0$.
	Therefore, (\ref{conservation}) means that operators in $\mathcal{P}$ are constant in time also in dNCQFT\footnote{To verify this statement, we must prove that the time evolution of operators in dNCQFT is given by the commutator with $H^\theta$. This can be easily seen by noticing that the time evolution of $\phi^\theta$ and $\pi^\theta$ is given by $[H^\theta,\phi^\theta]=i\dot{\phi}^\theta$ and $[H^\theta,\pi^\theta]=i\dot{\pi}^\theta$ respectively\cite{Abe:2006ig}.}:
\begin{equation}
\begin{array}{l}
	{\displaystyle \frac{dP_\mu}{dt}=\frac{1}{i}[H^\theta,P_\mu]=0,}\\
	{\displaystyle \frac{dM_{\mu\nu}}{dt}=\frac{\partial M_{\mu\nu}}{\partial t}+\frac{1}{i}[H^\theta,M_{\mu\nu}]=0.}\rule{0mm}{7mm}
\end{array}
\end{equation}

	For the representation space of $\mathcal{P}$ and $\mathcal{U}(\mathcal{P})$ represented by (\ref{rep. of poincare}) and (\ref{rep. in dNCQFT}), we can take the Hilbert space $\mathcal{H}$ and an algebra $\mathcal{M}(\phi^\theta)$ generated by products of the field operator $\phi^\theta$ in the same way as in CQFT.
	Notice that we can use the same Hilbert space $\mathcal{H}$ to represent the field operator for both CQFT and dNCQFT by based on the map (\ref{map}), as we mentioned in section 2.
	For brevity, we use the same symbols for each product maps in $\mathcal{H}$, $\mathcal{H}^*$ and $\mathcal{M}(\phi^\theta)$ as those corresponding maps introduced in section 3.1.
	That is,
\begin{equation}
\begin{array}{ccl}
	\mu&:&\mathcal{M}(\phi^\theta)\otimes\mathcal{M}(\phi^\theta)\rightarrow\mathcal{M}(\phi^\theta),\quad\mu(O_1^\theta\otimes O_2^\theta)=O_1^\theta O_2^\theta,\\
	\mu_R&:&\mathcal{M}(\phi^\theta)\otimes\mathcal{H}\rightarrow\mathcal{H},\quad\mu_R(O^\theta\otimes|\alpha\rangle)=O^\theta|\alpha\rangle,\rule{0mm}{5mm}\\
	\mu_L&:&\mathcal{H}^*\otimes\mathcal{M}(\phi^\theta)\rightarrow\mathcal{H}^*,\quad\mu_L(\langle\alpha|\otimes O^\theta)=\langle\alpha|O^\theta,\rule{0mm}{5mm}\\
	\widetilde{\rm ev}&:&\mathcal{H}^*\otimes\mathcal{M}(\phi^\theta)\otimes\mathcal{H}\rightarrow\mathbb{C},\quad\widetilde{\rm ev}(\langle\alpha|\otimes O^\theta\otimes|\beta\rangle)=\langle\alpha|O^\theta|\beta\rangle,\rule{0mm}{5mm}
\end{array}
	\label{product maps}
\end{equation}
	where $O^\theta,O_1^\theta,O_2^\theta\in\mathcal{M}(\phi^\theta)$.
	The Leibniz rule of the action of $\mathcal{U}(\mathcal{P})$ on product maps, (\ref{action to inner product}), (\ref{action to product for operators}), (\ref{action to state with operator}) and (\ref{action to matrix element}), and relations between product maps such as (\ref{map for matrix element}) and (\ref{associativity for operators and state}) which we have seen in the previous subsection also hold for these product.
	In particular, from the relation corresponding to (\ref{action to product for operators}) we can see that $\mathcal{M}(\phi^\theta)$ represents $\mathcal{U}(\mathcal{P})$ as a module algebra.

	So far, there seems no difference between the representation of $\mathcal{U}(\mathcal{P})$ in CQFT and that of dNCQFT.
	The difference appears in the action of $\mathcal{U}(\mathcal{P})$ to $\mathcal{M}(\phi^\theta)$.
	For $P_\mu, M_{\mu\nu}\in\mathcal{P}\subset\mathcal{U}(\mathcal{P})$ and $\phi^\theta\in\mathcal{M}(\phi^\theta)$, this action is calculated through the representation (\ref{rep. of poincare}) and (\ref{rep. in dNCQFT}), and the commutation relation (\ref{deformed quantization}):
\begin{equation}
\begin{array}{ccl}
	P_\mu(\phi^\theta)&:=&[P_\mu,\phi^\theta]=-i\partial_\mu\phi^\theta,\\
	M_{\mu\nu}(\phi^\theta)&:=&[M_{\mu\nu},\phi^\theta]\rule{0mm}{7mm}\\
	&=&-i(x_\mu\partial_\nu-x_\nu\partial_\mu)\phi^\theta-{\displaystyle \frac{i}{2}\Big[{\theta_\mu}^i(P_i\delta_\nu^\alpha-P_\nu\delta_i^\alpha)-{\theta_\nu}^i(P_i\delta_\mu^\alpha-P_\mu\delta_i^\alpha)\Big]\partial_\alpha\phi^\theta}.\rule{0mm}{5mm}
\end{array}
	\label{twisted action of poincare}
\end{equation}
	Notice that the action of generators of a Lorentz transformation $M_{\mu\nu}$ to $\phi^\theta$ is different from the standard one (\ref{action of poincare}).
	This action is exponentiated to give a finite Lorentz transformation ${\Lambda^\mu}_\nu$. This finite Lorentz transformation of $\phi^\theta$ can be written formally as
$$
\begin{CD}
	\phi^\theta(x^\mu)@>>\Lambda >\phi^\theta({\Lambda^\mu}_\nu x^\nu+\frac{1}{2}{\Lambda^\mu}_\nu\theta^{\nu\rho}P_\rho-\frac{1}{2}\theta^{\mu\nu}{\Lambda_\nu}^\rho P_\rho).
\end{CD}
$$
	The change of the coordinate induced by the Lorentz transformation has the form similar to the noncommutative Lorentz transformation in \cite{Calmet:2004ii}. In fact, it is considered as the field theoretical expression of the noncommutative Lorentz transformation in \cite{Calmet:2004ii}. For the case of free field, this result is consistent with \cite{Joung:2007qv}.

	Now that the structure of $\mathcal{P}$ or $\mathcal{U}(\mathcal{P})$ represented on dNCQFT is clarified, we twist $\mathcal{U}(\mathcal{P})$ and its representation spaces.
	By twisting $\mathcal{U}(\mathcal{P})$ by the invertible element $\mathcal{F}=e^{\frac{i}{2}\theta^{ij}P_i\otimes P_j}$, we obtain the twisted Poincar\'e algebra $\mathcal{U}^\mathcal{F}(\mathcal{P})$ which has the following coalgebraic structure:
\begin{equation}
\begin{array}{rcl}
	\Delta^\mathcal{F}(X^t)&=&\mathcal{F}\Delta(X)\mathcal{F}^{-1},\\
	\varepsilon^\mathcal{F}(X^t)&=&\varepsilon(X),\rule{0mm}{5mm}\\
	S^\mathcal{F}(X^t)&=&S(X),\rule{0mm}{5mm}
\end{array}
\end{equation}
	where $X^t\in\mathcal{U}^\mathcal{F}(\mathcal{P})$ is the same element as $X\in\mathcal{U}(\mathcal{P})$ as an element of the algebra.
	For $P_\mu^t,M_{\mu\nu}^t\in\mathcal{P}\subset\mathcal{U}^\mathcal{F}(\mathcal{P})$, this coproduct gives
\begin{equation}
\begin{array}{ccl}
	\Delta^\mathcal{F}(P_\mu^t)&=&P_\mu^t\otimes 1+1\otimes P_\mu^t,\\
	\Delta^\mathcal{F}(M_{\mu\nu}^t)&=&M_{\mu\nu}^t\otimes 1+1\otimes M_{\mu\nu}^t\rule{0mm}{5mm}\\
	&&{\displaystyle -\frac{1}{2}\theta^{ij}\Big[(g_{i\mu}P_\nu^t-g_{i\nu}P_\mu^t)\otimes P_j^t+P_i^t\otimes(g_{j\mu}P_\nu^t-g_{j\nu}P_\mu^t)\Big]}.\rule{0mm}{5mm}
\end{array}
\end{equation}
	The procedure of the twist induces the way for deriving a module algebra of $\mathcal{U}^\mathcal{F}(\mathcal{P})$ from a module algebra of $\mathcal{U}(\mathcal{P})$.
	In the case of $\mathcal{M}(\phi^\theta)$, by twisting the product map $\mu:\mathcal{M}(\phi^\theta)\otimes\mathcal{M}(\phi^\theta)\rightarrow\mathcal{M}(\phi^\theta)$ as
\begin{equation}
\begin{array}{rcl}
	\mu^\mathcal{F}(O_1^\theta\otimes O_2^\theta)&:=&\mu(\mathcal{F}^{-1}(O_1^\theta\otimes O_2^\theta)),\\
	&=:&O_1^\theta\star O_2^\theta,\rule{0mm}{5mm}
\end{array}
\end{equation}
	we obtain a module algebra $\mathcal{M}^\mathcal{F}(\phi^\theta)$ of $\mathcal{U}^\mathcal{F}(\mathcal{P})$.
	That is, the algebra $\mathcal{M}^\mathcal{F}(\phi^\theta)$ generated from products of field operators $\phi^\theta$ with the product map $\mu^\mathcal{F}$ gives a module algebra of $\mathcal{U}^\mathcal{F}(\mathcal{P})$.
	Here we use the same symbol $\star$ for this product as the extended star product (\ref{star product}).
	It is easily seen that, for field operators $\phi^\theta(x)$ and $\phi^\theta(y)$, this product gives the extended star product (\ref{star product}):
\begin{equation}
	\mu^\mathcal{F}(\phi^\theta(x)\otimes\phi^\theta(y))=\phi^\theta(x)\star\phi^\theta(y).
\end{equation}
	In addition to $\mu$, we introduce a twisted product for other product maps by the same procedure.
	First, we twist the map for the inner product of $\mathcal{H}$ (\ref{inner product}),
\begin{equation}
	{\rm ev}^\mathcal{F}(\langle\alpha|\otimes|\beta\rangle):={\rm ev}(\mathcal{F}^{-1}(\langle\alpha|\otimes|\beta\rangle))=:\langle\alpha|\star|\beta\rangle.\label{twisted inner product}
\end{equation}
	This seems to provide a new inner product for the Hilbert space $\mathcal{H}$, but in fact, ${\rm ev}^\mathcal{F}={\rm ev}$ since
\begin{equation}
	\langle\alpha|\star|\beta\rangle=\langle\alpha|\exp\Big[\frac{i}{2}P_i\theta^{ij}P_j\Big]|\beta\rangle=\langle\alpha|\beta\rangle.
\end{equation}
	We insert $\star$ in the inner product only to make explicit the associativity of products in calculating matrix elements, as we shall see below.
	We also introduce a star products for actions of $\mathcal{M}^\mathcal{F}(\phi^\theta)$ to $\mathcal{H}$ and $\mathcal{H}^*$,
\begin{equation}
\begin{array}{rcl}
	\mu_R^\mathcal{F}&:&\mathcal{M}^\mathcal{F}(\phi^\theta)\otimes\mathcal{H}\rightarrow\mathcal{H},\\
	&&\mu_R^\mathcal{F}(O^\theta\otimes|\alpha\rangle):=\mu_R(\mathcal{F}^{-1}(O^\theta\otimes |\alpha\rangle))=:O^\theta\star|\alpha\rangle,\rule{0mm}{5mm}\\
	\mu_L^\mathcal{F}&:&\mathcal{H}^*\otimes\mathcal{M}^\mathcal{F}(\phi^\theta)\rightarrow\mathcal{H}^*,\rule{0mm}{5mm}\\
	&&\mu_L^\mathcal{F}(\langle\alpha|\otimes O^\theta):=\mu_L(\mathcal{F}^{-1}(\langle\alpha|\otimes O^\theta))=:\langle\alpha|\star O^\theta.\rule{0mm}{5mm}
\end{array}
\end{equation}
	Finally we introduce a linear map $\widetilde{\rm ev}^\mathcal{F}:\mathcal{H}^*\otimes\mathcal{M}^\mathcal{F}(\phi^\theta)\otimes\mathcal{H}\rightarrow\mathbb{C}$ for evaluating matrix elements of operators in $\mathcal{M}^\mathcal{F}(\phi^\theta)$,
\begin{equation}
\begin{array}{rcl}
	\widetilde{\rm ev}^\mathcal{F}(\langle\alpha|\otimes O^\theta\otimes|\beta\rangle)&:=&{\rm ev}^\mathcal{F}\circ(1\otimes\mu_R^\mathcal{F})(\langle\alpha|\otimes O^\theta\otimes|\beta\rangle)\\
	&=&{\rm ev}^\mathcal{F}\circ(\mu_L^\mathcal{F}\otimes 1)(\langle\alpha|\otimes O^\theta\otimes|\beta\rangle)\rule{0mm}{5mm}\\
	&=:&\langle\alpha|\star O^\theta\star|\beta\rangle.\rule{0mm}{5mm}
\end{array}
	\label{twisted matrix element}
\end{equation}
	The second equality of this equation means $\langle\alpha|\star(O^\theta\star|\beta\rangle)=(\langle\alpha|\star O^\theta)\star|\beta\rangle$, i.e., associativity of the star product.
	This can be easily proved.
	In fact, noticing
\begin{equation}
	(1\otimes\Delta)(\mathcal{F}^{-1})(1\otimes\mathcal{F}^{-1})=e^{-\frac{i}{2}\theta^{ij}(P_i\otimes P_j\otimes 1+P_i\otimes 1\otimes P_j+1\otimes P_i\otimes P_j)}=(\Delta\otimes 1)(\mathcal{F}^{-1})(\mathcal{F}^{-1}\otimes 1),\label{associativity of star product}
\end{equation}
	we find
\begin{equation}
\begin{array}{rcccl}
	{\rm ev}^\mathcal{F}\circ(1\otimes\mu_R^\mathcal{F})&=&{\rm ev}\circ(1\otimes\mu_R)\circ(1\otimes\Delta)(\mathcal{F}^{-1})(1\otimes\mathcal{F}^{-1})&&\\
	&=&{\rm ev}\circ(\mu_L\otimes 1)\circ(\Delta\otimes 1)(\mathcal{F}^{-1})(\mathcal{F}^{-1}\otimes 1)&=&{\rm ev}^\mathcal{F}\circ(\mu_L^\mathcal{F}\otimes 1),\rule{0mm}{5mm}
\end{array}
	\label{associativity of matrix elements}
\end{equation}
	where we use (\ref{map for matrix element}).
	This proof is essentially the same as the proof of associativity of the ordinary Moyal star product, which also uses (\ref{associativity of star product}).
	Furthermore, we can show associativity for all the star products introduced here in the same way.
	For example, quantities such as
\begin{equation}
	O_1^\theta\star O_2^\theta\star O_3^\theta\star|\alpha\rangle,\quad\langle\alpha|\star O_1^\theta\star O_2^\theta\star|\beta\rangle,
\end{equation}
	do not depend on an order of taking products in them.

	Next, we observe a relation between these star products and a twisted Poincar\'e transformation.
	In the first place, since, as we noted above, the algebra $\mathcal{M}^\mathcal{F}(\phi^\theta)$ is a module algebra of $\mathcal{U}^\mathcal{F}(\mathcal{P})$, a twisted Poincar\'e transformation of the star product of $\mathcal{M}^\mathcal{F}(\phi^\theta)$ is given by
\begin{equation}
\begin{array}{rcl}
	X^t(\phi^\theta(x)\star\phi^\theta(y))&=&X^t(\mu^\mathcal{F}(\phi^\theta(x)\otimes\phi^\theta(y)))\\
	&=&\mu^\mathcal{F}(\Delta^\mathcal{F}(X^t)(\phi^\theta(x)\otimes\phi^\theta(y))),\quad \textrm{for $X^t\in\mathcal{U}^\mathcal{F}(\mathcal{P})$.}\rule{0mm}{5mm}
\end{array}
\end{equation}
	For a twisted Poincar\'e transformation of other star products, we can verify the twisted Leibniz rule in the same form:
\begin{equation}
\begin{array}{l}
	X^t(\langle\alpha|\star|\beta\rangle)=X^t({\rm ev}^\mathcal{F}(\langle\alpha|\otimes|\beta\rangle))={\rm ev}^\mathcal{F}(\Delta^\mathcal{F}(X^t)(\langle\alpha|\otimes|\beta\rangle)),\rule{0mm}{5mm}\\
	X^t(O^\theta\star|\alpha\rangle)=X^t(\mu_R^\mathcal{F}(O^\theta\otimes|\beta\rangle))=\mu_R^\mathcal{F}(\Delta^\mathcal{F}(X^t)(O^\theta\otimes|\beta\rangle)),\rule{0mm}{5mm}\\
	X^t(\langle\alpha|\star O^\theta)=X^t(\mu_L^\mathcal{F}(\langle\alpha|\otimes O^\theta))=\mu_L^\mathcal{F}(\Delta^\mathcal{F}(X^t)(\langle\alpha|\otimes O^\theta)),\rule{0mm}{5mm}\\
\end{array}
	\label{twisted action to twisted products}
\end{equation}
	and using these relations and (\ref{associativity of matrix elements}), we obtain
\begin{equation}
\begin{array}{rcl}
	X^t(\langle\alpha|\star O^\theta\star|\beta\rangle)&=&X^t(\widetilde{\rm ev}^\mathcal{F}(\langle\alpha|\otimes O^\theta\otimes|\beta\rangle))\\
	&=&\widetilde{\rm ev}^\mathcal{F}((1\otimes\Delta^\mathcal{F})\circ\Delta^\mathcal{F}(X^t)(\langle\alpha|\otimes O^\theta\otimes|\beta\rangle))\rule{0mm}{5mm}\\
	&=&\widetilde{\rm ev}^\mathcal{F}((\Delta^\mathcal{F}\otimes 1)\circ\Delta^\mathcal{F}(X^t)(\langle\alpha|\otimes O^\theta\otimes|\beta\rangle)).\rule{0mm}{5mm}
\end{array}
	\label{twisted action to twisted matrix element}
\end{equation}

	Finally, we note that the inner product (\ref{twisted inner product}) is invariant under a twisted Poincar\'e transformation, as the inner product in CQFT (\ref{inner product}) is invariant under a Poincar\'e transformation.
	In fact, using a formula of an antipode of $\mathcal{U}^\mathcal{F}(\mathcal{P})$ which corresponds to (\ref{formula for antipode}),
\begin{equation}
	{\rm m}((S^\mathcal{F}\otimes 1)\circ\Delta^\mathcal{F}(X^t))=\varepsilon^\mathcal{F}(X^t)\Big(={\rm m}((1\otimes S^\mathcal{F})\circ\Delta^\mathcal{F}(X^t))\Big),
\end{equation}
	the action of a twisted Poincar\'e transformation to an inner product (i.e., the first equation in (\ref{twisted action to twisted products})) is written as
\begin{equation}
	X^t(\langle\alpha|\star|\beta\rangle)=\langle\alpha|\varepsilon^\mathcal{F}(X^t)|\beta\rangle.\label{invariance of star inner product}
\end{equation}
	Then, from the explicit value of the counit $\varepsilon^\mathcal{F}(X^t)=\varepsilon(X)$, (see (\ref{counit})) we find the invariance of the inner product $\langle\alpha|\star|\beta\rangle$.
	Furthermore, from this result and (\ref{twisted matrix element}), we can show that a matrix element of operators in dNCQFT is also invariant under a twisted Poincar\'e transformation.
\begin{equation}
	X^t(\langle\alpha|\star O^\theta\star|\beta\rangle)=\langle\alpha|\varepsilon^\mathcal{F}(X^t)(O^\theta\star|\beta\rangle)=(\langle\alpha|\star O^\theta)\varepsilon^\mathcal{F}(X^t)|\beta\rangle.\label{twisted invariance of ME}
\end{equation}

\section{Correspondence between the symmetries}\mbox{}

	In section 2, we have seen the correspondence between CQFT and dNCQFT established by (\ref{map}).
	In this section, we shall prove that this correspondence leads to the correspondence between the Poincar\'e symmetry of CQFT and the twisted Poincar\'e symmetry of dNCQFT.
	This statement is precisely expressed in the following theorem:
\vspace{5mm}

	\noindent {\bf Theorem 1.} {\it Let $O(\phi^0)\in\mathcal{M}(\phi^0)$ and $O_\star(\phi^\theta)\in\mathcal{M}^\mathcal{F}(\phi^\theta)$ be operators related with each other by (\ref{map}) and (\ref{operator correspondence}), and $|\alpha\rangle$ be an arbitrary state in the Hilbert space $\mathcal{H}$ on which $\phi^0$ and $\phi^\theta$ are represented.
	Then we have
\begin{equation}
	O_\star(\phi^\theta)\star|\alpha\rangle=O(\phi^0)|\alpha\rangle.\label{state correspondence}
\end{equation}
	Further, this equality holds when one transforms the left hand side by $X^t\in\mathcal{U}^\mathcal{F}(\mathcal{P})$, and right hand side by $X\in\mathcal{U}(\mathcal{P})$ where $X^t$ is the same element as $X$ as an element of the algebra:
\begin{equation}
	X^t(O_\star(\phi^\theta)\star|\alpha\rangle)=X(O(\phi^0)|\alpha\rangle),
\end{equation}
	or equivalently}
\begin{equation}
\begin{array}{rl}
	&X^t\Big(\mu_R^\mathcal{F}(O_\star(\phi^\theta)\otimes|\alpha\rangle)\Big)=X\Big(\mu_R(O(\phi^0)\otimes|\alpha\rangle)\Big)\\
	\Leftrightarrow&\mu_R^\mathcal{F}\Big(\Delta^\mathcal{F}(X^t)(O_\star(\phi^\theta)\otimes|\alpha\rangle)\Big)=\mu_R\Big(\Delta(X)(O(\phi^0)\otimes|\alpha\rangle)\Big).\rule{0mm}{5mm}
\end{array}
	\label{symmetry correspondence}
\end{equation}
\vspace{3mm}

	\noindent{\it Proof.} It is trivial to prove the first part of this theorem, i.e., (\ref{state correspondence}): substituting (\ref{operator correspondence}) into the left hand side of (\ref{state correspondence}), we immediately obtain the right hand side.
	To prove the second part, we introduce the following notation for the twisting element $\mathcal{F}$:
\begin{equation}
	\mathcal{F}=\sum_if'_i\otimes f''_i\in\mathcal{U}(\mathcal{P})\otimes\mathcal{U}(\mathcal{P}).
\end{equation}
	Using this notation, correspondences between fields (\ref{map}) and between operators (\ref{operator correspondence}) are rewritten as
\begin{equation}
	\phi^\theta=\sum_if'_i(\phi^0)f''_i,\quad O_\star(\phi^\theta)=\sum_if'_i(O(\phi^0))f''_i,\label{operator correspondence2}
\end{equation}
	where $f''_i$ is considered as an element not in $\mathcal{U}^\mathcal{F}(\mathcal{P})$ but in $\mathcal{M}^\mathcal{F}(\phi^\theta)$.
	In this notation, the inverse $\mathcal{F}^{-1}$ is given by
\begin{equation}
	\mathcal{F}^{-1}=\sum_if''_i\otimes f'_i,
\end{equation}
	and thus $\mathcal{F}\cdot\mathcal{F}^{-1}=\mathcal{F}^{-1}\cdot\mathcal{F}=1\otimes 1$ reads
\begin{equation}
	\sum_{i,j}f'_if''_j\otimes f''_if'_j=\sum_{i,j}f''_jf'_i\otimes f'_jf''_i=1\otimes 1.
\end{equation}
	Since $f'_i$ and $f''_i$ are given by the form of a polynomial of $P_i$ and commutative each other, we see further
\begin{equation}
	\sum_{i,j}f'_if''_j\otimes f'_jf''_i=\sum_{i,j}f''_jf'_i\otimes f''_if'_j=1\otimes 1.\label{identity of F}
\end{equation}

	To prove (\ref{symmetry correspondence}), we first show the following relation:
\begin{equation}
	\mathcal{F}^{-1}\Big(\sum_if'_i(O(\phi^0))f''_i\otimes|\alpha\rangle\Big)=\sum_{i,j}(\mu\otimes 1)(f''_jf'_i(O(\phi^0))\otimes f''_i\otimes f'_j|\alpha\rangle).\label{lemma}
\end{equation}
	For this purpose, we write $\mathcal{F}^{-1}$ in the equation explicitly by $P_i$:
\begin{equation}
	\textrm{L.H.S of (\ref{lemma})}=\sum_{i,n}\left(\frac{i}{2}\right)^n\frac{1}{n!}\theta^{i_1j_1}\cdots\theta^{i_nj_n}P_{i_1}\cdots P_{i_n}\Big(f'_i(O(\phi^0))f''_i\Big)\otimes P_{j_1}\cdots P_{j_n}|\alpha\rangle.\label{eq1 to lemma}
\end{equation}
	Since $f''_i$ is given by a form of a polynomial of $P_i$ and therefore commutes with $P_i$,
\begin{equation}
\begin{array}{rcl}
	P_{i_1}\cdots P_{i_n}\Big(f'_i(O(\phi^0))f''_i\Big)&=&\big[P_{i_1},\big[P_{i_2},\cdots\big[P_{i_n},f'_i(O(\phi^0))f''_i\big]\cdots\big]\big]\\
	&=&\big[P_{i_1},\big[P_{i_2},\cdots\big[P_{i_n},f'_i(O(\phi^0))\big]\cdots\big]\big]f''_i\rule{0mm}{5mm}\\
	&=&\Big(P_{i_1}\cdots P_{i_n}\big(f'_i(O(\phi^0))\big)\Big)f''_i\rule{0mm}{5mm}\\
	&=&\big(P_{i_1}\cdots P_{i_n}f'_i(O(\phi^0))\big)f''_i.\rule{0mm}{5mm}
\end{array}
\end{equation}
	Then (\ref{eq1 to lemma}) reads
\begin{equation}
\begin{array}{rcl}
	(\textrm{\ref{eq1 to lemma}})&=&\displaystyle{\sum_{i,n}\left(\frac{i}{2}\right)^n\frac{1}{n!}\theta^{i_1j_1}\cdots\theta^{i_nj_n}\big(P_{i_1}\cdots P_{i_n}f'_i(O(\phi^0))\big)f''_i\otimes P_{j_1}\cdots P_{j_n}|\alpha\rangle}\\
	&=&\displaystyle{\sum_{i,j}\big(f''_jf'_i(O(\phi^0))\big)f''_i\otimes f'_j|\alpha\rangle}\rule{0mm}{5mm}\\
	&=&{\displaystyle \sum_{i,j}(\mu\otimes 1)\big(f''_jf'_i(O(\phi^0))\otimes f''_i\otimes f'_j|\alpha\rangle\big)}\rule{0mm}{5mm}
\end{array}
\end{equation}
	and this is just the right hand side of (\ref{lemma}).

	Using (\ref{operator correspondence2}) and (\ref{lemma}), the left hand side of (\ref{symmetry correspondence}) reads
\begin{equation}
\begin{array}{rcl}
	\mu_R^\mathcal{F}\Big(\Delta^\mathcal{F}(X^t)(O_\star(\phi^\theta)\otimes|\alpha\rangle)\Big)&=&{\displaystyle \sum_{i,j}\mu_R\Big(\Delta(X)(\mu\otimes 1)(f''_jf'_i(O(\phi^0))\otimes f''_i\otimes f'_j|\alpha\rangle)\Big)}\\
	&=&{\displaystyle \sum_{i,j}\mu_R\circ(\mu\otimes 1)\Big((\Delta\otimes 1)\circ\Delta(X)(f''_jf'_i(O(\phi^0))\otimes f''_i\otimes f'_j|\alpha\rangle)\Big)}\\
	&=&{\displaystyle \sum_{i,j}\mu_R\circ(1\otimes \mu_R)\Big((1\otimes\Delta)\circ\Delta(X)(f''_jf'_i(O(\phi^0))\otimes f''_i\otimes f'_j|\alpha\rangle)\Big)}\\
	&=&{\displaystyle \sum_{i,j}X\Big(\mu_R\circ(1\otimes\mu_R)(f''_jf'_i(O(\phi^0))\otimes f''_i\otimes f'_j|\alpha\rangle)\Big)}\\
	&=&{\displaystyle \sum_{i,j}X\Big(\mu_R(f''_jf'_i(O(\phi^0))\otimes f''_if'_j|\alpha\rangle)\Big).}
\end{array}
	\label{equation}
\end{equation}
	To show the third equality, we use coassociativity of the coproduct $\Delta$ and associativity of products $\mu_R$ and $\mu$ (\ref{associativity for operators and state}).
	Finally, by using (\ref{identity of F}), we obtain the right hand side of (\ref{symmetry correspondence}).
	That is,
\begin{equation}
\begin{array}{rcll}
	\textrm{R.H.S of (\ref{equation})}&=&{\displaystyle \sum_{i,j}X\Big(\mu_R((f''_jf'_i\otimes f''_if'_j)(O(\phi^0)\otimes|\alpha\rangle))\Big)}&\\
	&=&X\Big(\mu_R(O(\phi^0)\otimes|\alpha\rangle)\Big).&\Box\rule{0mm}{5mm}
\end{array}
\end{equation}
\vspace{5mm}

	For completeness, we prove the correspondence between a Poincar\'e transformation of the inner product and matrix elements in CQFT and a twisted Poincar\'e transformation of them in dNCQFT.\vspace{5mm}

	\noindent{\bf Theorem 2.} {\it Let $O(\phi^0)$, $O_\star(\phi^\theta)$, $X^t$ and $X$ be as in Theorem 4.1, and $\langle\alpha|$ and $|\beta\rangle$ be arbitrary elements in $\mathcal{H}^\ast$ and $\mathcal{H}$ respectively.
	Then we have
\begin{equation}
	X^t(\langle\alpha|\star|\beta\rangle)=X(\langle\alpha|\beta\rangle),\label{correspondence of inner product}
\end{equation}
	and}
\begin{equation}
	X^t(\langle\alpha|\star O_\star(\phi^\theta)\star|\beta\rangle)=X(\langle\alpha|O(\phi^0)|\beta\rangle.\label{correspondence of matrix element}
\end{equation}
\vspace{3mm}

	\noindent{\it Proof.} Since (\ref{correspondence of inner product}) is given by the case where $O(\phi^0)=O_\star(\phi^\theta)=1$ in (\ref{correspondence of matrix element}), it is suffice to prove (\ref{correspondence of matrix element}).
	This is easily done by using (\ref{invariance of matrix elements}) and (\ref{twisted invariance of ME}):
\begin{equation}
\begin{array}{rclcl}
	X(\langle\alpha|O(\phi^0)|\beta\rangle)&=&\langle\alpha|\varepsilon(X)\big(O(\phi^0)|\beta\rangle\big)&&\\
	&=&\langle\alpha|\varepsilon^\mathcal{F}(X^t)\big(O_\star(\phi^\theta)\star|\beta\rangle\big)&=&X^t\big(\langle\alpha|\star O_\star(\phi^\theta)\star|\beta\rangle\big).\rule{0mm}{6mm}
\end{array}
	\label{symmetry correspondence2}
\end{equation}
	where we use (\ref{state correspondence}) to prove second equality.\hspace{1cm}$\Box$
\vspace{5mm}

	From the results obtained here, in particular (\ref{symmetry correspondence}) and (\ref{symmetry correspondence2}), one can see that the Poincar\'{e} covariance of CQFT implies the twisted Poincar\'e covariance of dNCQFT.
	Thus, dNCQFT gives an example of a QFT whose symmetry is described by a quantum group.
	If the symmetry group of dNCQFT is restricted to a classical group, it is given by a reduced Poincar\'e group, e.g., in the case of four dimensional space-time, the symmetry group is $[O(1,1)\times SO(2)]\rtimes \mathcal{T}_4$.

\section{Conclusions and remarks}\mbox{}

	We have discussed the twisted Poincar\'e symmetry of noncommutative QFTs with the deformed quantization (dNCQFT) and their correspondence with the Poincar\'e symmetry of standard commutative QFTs (CQFT).
	We have seen that the equivalence in correlation functions between dNCQFT and CQFT is established by the map (\ref{map}) and have presented the rigorous proof of the correspondence between symmetries of the two theories.
	By use of the map, we can represent generators of the twisted Poincar\'e algebra by operators acting on a Hilbert space on which the field operator of dNCQFT is represented.
	It is easy to see that a twisted Poincar\'e transformation on dNCQFT constructed in this way is translated to a Poincar\'e transformation on CQFT by the aid of the map between the two theories.
	This result is seemingly surprising: the two different types of symmetries correspond with each other through the QFTs with different types of quantization schemes.
	We see that actually, this correspondence is made clear by presenting both symmetries in terms of a Hopf algebra.
	From a Hopf algebraic point of view, both the Poincar\'e algebra and the twisted Poincar\'e algebra are quantum groups and the only difference is that the former is cocommutative while the latter is noncocommutative.

	In the process of constructing the twisted Poincar\'e algebra, we obtain a conserved charge associated to the transformation.
	This is essential to our analysis, since without such operators constant in time, it would be difficult to construct the Poincar\'e algebra in dNCQFT and represent the twisted Poincar\'e transformation on the dNCQFT as an operator acting in the Hilbert space.
	Indeed, it has not been obtained by simple application of the Noether procedure extended to the case of the twisted Poincar\'e algebra\cite{Gonera:2005hg}.

	In this paper, we have proved the correspondence between symmetries of CQFT and dNCQFT underlying the equivalence between the two theories.
	We have mentioned in \cite{Abe:2006ig} that the equivalence of correlation functions may be seen for more general theories.
	In fact, if we use different noncommutative parameters for the interaction term in (\ref{noncommutative lagrangian}) and for commutator in (\ref{deformed quantization}), say $\theta^{ij}$ and $\tilde\theta^{ij}$, respectively, then the resulting dynamics of the theory depends only on the difference between them $\Theta^{ij}=\theta^{ij}-\tilde\theta^{ij}$.
	In particular, all the deformed QFTs which have the same value of $\Theta^{ij}$ are equivalent to the ordinary noncommutative QFT with the noncommutative parameter $\Theta^{ij}$ in their dynamics.
	This suggests that all the twisted Poincar\'e symmetries in theories sharing the same $\Theta^{ij}$ would also correspond each other in their generators and coproduct through a map establishing the equivalence.
	However, we cannot apply the method employed here straightforwardly to prove this general correspondence of symmetries, because it is not clear how to construct an operator associated to a twisted Poincar\'e transformation from the field operator in the noncommutative QFT (or in general dNCQFTs) by the same procedure.
	This prevent us from representing the twisted Poincar\'e algebra on the Hilbert space carrying the representation of the field operator of the theory.
	In other words, the situation becomes especially simple in the case $\Theta^{ij}=0$ which we have considered in this paper.
	The specialty of the case $\Theta^{ij}=0$ would be expected from the fact that, at least in classical level, Moyal star products with the same rank but different value of $\theta^{ij}$ give rise to Morita equivalent algebras.
	The difference between the property of $\Theta^{ij}=0$ theory and that of $\Theta^{ij}\ne0$ theory would reflect this equivalence in classical level.
	Despite the difficulty in the extension, however, we believe that the result and the method presented in this paper provide a clue to a fuller understanding of the symmetry of the dNCQFT.
\newline

\noindent{\bf \large{Acknowledgments}}\mbox{}\newline

	I would like to thank Izumi Tsutsui for many useful discussions and a careful reading of the manuscript.

\end{document}